# 高級加密標準的各種模式效能比較

# Performance Comparison of Various Modes of Advanced Encryption Standard

Abel C. H. Chen
Chunghwa Telecom Laboratories
ORCID: 0000-0003-3628-3033

## 摘要

隨著量子計算技術的成熟，許多密碼學方法逐漸面臨量子計算的威脅。儘管 Grover 演算法可以加速搜索速度，但目前研究成果表明高級加密標準(Advanced Encryption Standard, AES)方法仍然可以通過增加秘密金鑰(Secret Key)的長度來增強安全性。然而，高級加密標準方法在實作涉及多種模式(Mode)，並且非所有模式都是安全的。因此，本研究提出正規化 Gini 不純度(Normalized Gini Impurity, NGI)的評估指標來驗證每種模式的安全性，並且採用加密影像作為案例研究進行實證分析。此外，本研究主要比較電子密碼本(Electronic Codebook, ECB)模式、密文區塊鏈接(Cipher Block Chaining, CBC)模式、計數器(Counter, CTR)模式、帶密文區塊鏈接-訊息驗證碼計數器(Counter with CBC-Message Authentication Code (MAC), CCM)模式、以及 Galois 計數器模式(Galois Counter Mode, GCM)。

**關鍵詞**：高級加密標準、Gini 不純度、密文區塊鏈接模式、計數器模式。

## Abstract

With the maturation of quantum computing technology, many cryptographic methods are gradually facing threats from quantum computing. Although the Grover algorithm can accelerate search speeds, current research indicates that the Advanced Encryption Standard (AES) method can still enhance security by increasing the length of the secret key. However, the AES method involves multiple modes in implementation, and not all modes are secure. Therefore, this study proposes a normalized Gini impurity (NGI) to verify the security of each mode, using encrypted images as a case study for empirical analysis. Furthermore, this study primarily compares the Electronic Codebook (ECB) mode, Cipher Block Chaining (CBC) mode, Counter (CTR) mode, Counter with CBC-Message Authentication Code (MAC) (CCM) mode, and Galois Counter Mode (GCM).

# 一、前言

近年來，隨著量子計算和 Shor 量子演算法的發展，將可能快速解決質因數分解問題和離散對數問題[1]，導致非對稱式密碼學(如：RSA 密碼學和橢圓曲線密碼學)逐漸面臨量子計算的威脅。雖然對稱式密碼學—高級加密標準(Advanced Encryption Standard, AES)方法仍然具備抵抗量子計算攻擊的能力，可以通過秘密金鑰(Secret Key)的長度來抵禦 Grover 量子演算法的快速搜索[2]。然而，高級加密標準方法的安全性將取決於其模式(Mode)，部分模式可能存在安全風險。

有鑑於高級加密標準方法各種模式的安全風險，本研究主要深入探討高級加密標準方法的各種模式，比較它們的優勢和劣勢。主流高級加密標準方法的模式包括電子密碼本(Electronic Codebook, ECB)模式[3]、密文區塊鏈接(Cipher Block Chaining, CBC)模式[4]、計數器(Counter, CTR)模式[5]、帶密文區塊鏈接-訊息驗證碼計數器(Counter with CBC-Message Authentication Code (MAC), CCM)模式[6]、以及 Galois 計數器模式(Galois Counter Mode, GCM)[7]。

本研究主要提出評估指標來評估高級加密標準方法每個模式的安全性，然後比較每個模式的加密和解密時間。本研究的貢獻如下：

- 提出正規化 Gini 不純度[8]的評估指標，用於分析高級加密標準方法每個模式的安全性。
- 對各種高級加密標準方法模式的效能進行比較分析，包括電子密碼本模式、密文區塊鏈接模式、計數器模式、帶密文區塊鏈接-訊息驗證碼計數器模式、以及 Galois 計數器模式。
- 比較不同秘密金鑰長度下的效能，如：AES-128 [9]、AES-192 [10]、以及 AES-256 [11]。

本文共有五節。第二節介紹高級加密標準方法各種模式。第三節提出評估指標，並對每個模式進行加密影像的安全性評估。第四節測量每個模式的加密時間和解密時間。最後，第五節總結本研究貢獻，並且討論未來可行的研究方向。

# 二、高級加密標準方法各種模式

本節全面介紹高級加密標準方法各種模式。為了清楚描述和比較各種模式，原始明文 $P$ 被分割成 $n$ 個區塊；其中，第 $i$ 個明文區塊表示為 $p_i$，對應的第 $i$ 個密文區塊表示為 $c_i$。高級加密標準方法的基本加密函數表示為 $f(k, b)$，具有秘密金鑰 $k$ 和輸入數據 $b$，進行 SubBytes、ShiftRows、MixColumns 和 AddRoundKey 等操作[12]。後續小節分別詳細介紹電子密碼本模式、密文區塊鏈接模式、計數器模式、帶密文區塊鏈接-訊息驗證碼計數器模式、以及 Galois 計數器模式。

## 1. 電子密碼本模式

在高級加密標準方法的電子密碼本模式[3]中，根據公式(1)計算第 $i$ 個密文區塊[12]。在電子密碼本模式下，密文區塊 $c_i$ 主要受到秘密金鑰 $k$ 和第 $i$ 個明文區塊 $p_i$ 的影響。此外，可以觀察到相同的明文區塊內容會產生相同的密文區塊。因此，當 $p_i$ 和 $p_j$ 相同時，$c_i$ 和 $c_j$ 將會相同。因此，即使高級加密標準方法的基本加密函數 $f(k, b)$ 本身是安全的，但高級加密標準方法的電子密碼本模式仍可能表現出相對較低的安全性。

$$c_i = f(k, p_i) \tag{1}$$

**2. 密文區塊鏈接模式**

在高級加密標準方法的密文區塊鏈接模式[4]中，產製隨機初始值(Initial Value, IV)，表示為 $c_0$，作為第一個密文區塊。此外，使用 XOR 運算符⊕來計算第 $i$ 個明文區塊 $p_i$ 和第($i$–1)個密文區塊 $c_{i-1}$，使用公式(2)來產製第 $i$ 個密文區塊建立區塊鏈[12]。因此，即使 $p_i$ 和 $p_j$ 相同，$c_i$ 和 $c_j$ 也可能不同。除非 $p_i$ 和 $p_j$ 相同且 $c_{i-1}$ 和 $c_{j-1}$ 也相同，$c_i$ 和 $c_j$ 才會相同。但大部分的情況下，不太可能會發生，所以高級加密標準方法的密文區塊鏈接模式相對安全。

$$c_i = f(k, p_i \oplus c_{i-1}) \tag{2}$$

**3. 計數器模式**

在高級加密標準方法的計數器模式[5]中，產製隨機數(nonce) r 和計數器 $d$，搭配公式(3)來產製第 $i$ 個密文區塊[12]。計數器模式可以確保每個區塊中使用的值($r \| (d + i - 1)$)都是不同的。因此，即使 $p_i$ 和 $p_j$ 相同，也會得到不同的 $c_i$ 和 $c_j$，所以計數器模式相對安全。除此之外，計數器模式與密文區塊鏈接模式在計算效率上大不同；由於在計數器模式中，第 $i$ 個密文區塊 $c_i$ 的計算不依賴於第($i$–1)個密文區塊 $c_{i-1}$ 的計算結果，所以每個密文區塊可以平行計算，從而更快地計算出每個密文區塊的值。

$$c_i = p_i \oplus f(k, r \| (d + i - 1)) \tag{3}$$

**4. 帶密文區塊鏈接-訊息驗證碼計數器模式**

為確保完整性，可以產製訊息驗證碼用以驗證訊息是否被篡改，從而提高安全性。帶密文區塊鏈接-訊息驗證碼計數器模式採用密文區塊鏈接模式來產製訊息驗證碼並作為標籤值，然後再使用計數器模式對明文進行加密，以及把標籤值附加到密文中[12]。因此，與電子密碼本模式、密文區塊鏈接模式和計數器模式相比，帶密文區塊鏈接-訊息驗證碼計數器模式還提供訊息完整性的驗證。然而，由於採用密文區塊鏈接模式來產製訊息驗證碼，所以需要較多的計算時間。

**5. Galois 計數器模式**

Galois 計數器模式是另一種可以做訊息完整性驗證的模式，它採用基於計數器模式的訊息驗證碼來驗證訊息完整性。在 Galois 計數器模式中，使用公式(4)來產製訊息驗證碼，其中 $h(f(k, 0), C)$ 代表雜湊函數，$C$ 表示密文[12]。在 Galois 計數器模式中，每個密文區塊都是使用計數器模式計算，然後與相關資料結合以產製訊息驗證碼。Galois 計數器模式的優點在於能夠產製用於驗證訊息完整性的訊息驗證碼，同時具備計數器模式的平行計算特性。

$$t = h(f(k, 0), C) \oplus f(k, r \| 0) \tag{4}$$

## 三、安全性評估

第 1 小節提出評估指標用以量化高級加密標準方法每種模式的安全指數。第 2 小節描述本

研究的實驗設置和使用的原始影像。第 3 小節展示每種模式產生的加密影像及其評估結果。

**1. 評估指標**

為了驗證每種模式加密結果的安全性，本研究主要比較明文和密文之間的差異。假設原始影像的位元陣列表示為 $P$，加密影像的位元陣列表示為 $C$，則兩個影像之間的差異 $d$ 可以使用公式(5)計算，然後使用公式(6)計算相異位元發生的機率 $p_1$；其中，$b_1(d)$ 表示位元陣列 $d$ 中設置為 1 的位元數量，而 $L$ 表示位元陣列 $d$ 的長度。

由於原始的 Gini 不純度(Gini Impurity)[8]並不會落在 0 到 1 的範圍內，所以本研究提出公式(7)建立正規化 Gini 不純度(Normalized Gini Impurity, NGI) $g$。當正規化 Gini 不純度越高，則表示具有較高的安全性。

$$d = P \oplus C \tag{5}$$
$$p_1 = b_1(d) / L \tag{6}$$
$$g = 2 - 2[p_1^2 + (1 - p_1)^2] \tag{7}$$

**2. 實驗環境**

在評估高級加密標準方法每種模式的安全性時，本研究採用正規化 Gini 不純度作為評估指標。此外，本研究採用圖 1 中的 9 張影像作為原始影像，並且使用高級加密標準方法的每種模式對原始影像加密產製加密影像。在實驗環境中的軟硬體規格描述如下：CPU Intel(R) Core(TM) i7-10510U、RAM 8 GB、OpenJDK 18.0.2.1、以及函式庫 Bouncy Castle Release 1.72。

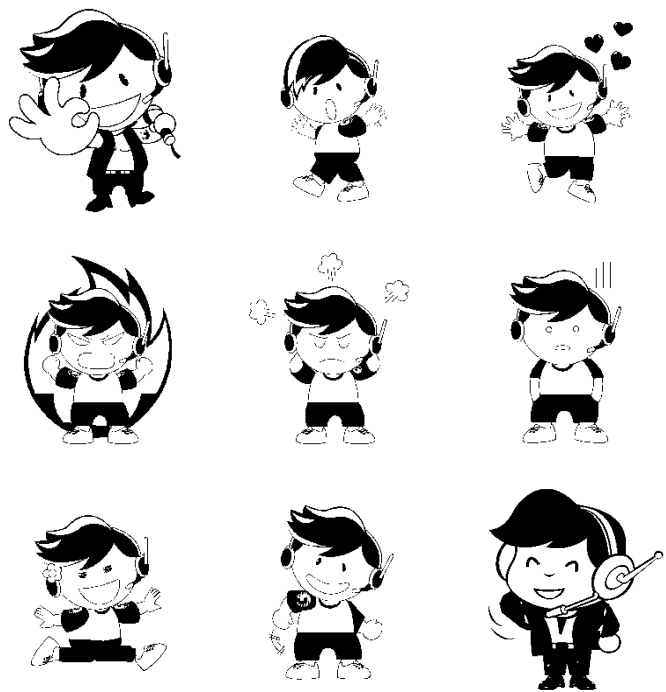

**圖 1 原始影像**

# 3. 實驗結果

本小節評估在不同秘密金鑰長度 AES-128、AES-192、以及 AES-256 的電子密碼本模式、密文區塊鏈接模式、計數器模式、帶密文區塊鏈接-訊息驗證碼計數器模式、以及 Galois 計數器模式的安全性。

## 3.1 AES-128實驗結果

圖2展示由 AES-128電子密碼本模式產製的加密影像。從實驗結果可以明顯看出，由電子密碼本模式產製的加密影像與原始影像相似。因此，電子密碼本模式的安全性相對較低。

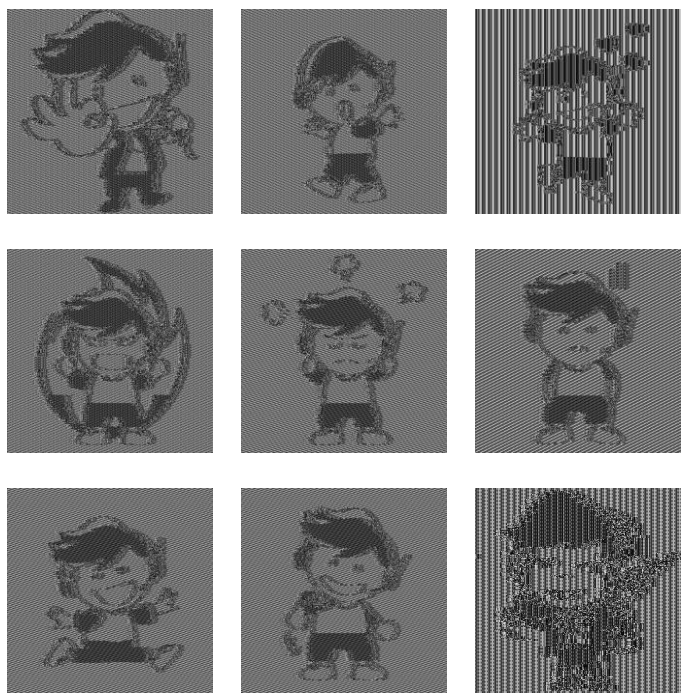

圖 2 AES-128電子密碼本模式產製的加密影像

圖3、圖4、圖5和圖6分別展示由 AES-128 密文區塊鏈接模式、計數器模式、帶密文區塊鏈接-訊息驗證碼計數器模式、以及 Galois 計數器模式產製的加密影像。從實驗結果觀察到，這些模式的加密影像與原始影像沒有直接可辨識的相似之處，表明這些模式的加密效果相對安全。

本研究使用提出的正規化 Gini 不純度來評估每種模式的安全性，如圖 7 所示。從實驗結果可以觀察到，電子密碼本模式具有最低的正規化 Gini 不純度，表明該模式的安全性最低。此外，可以觀察到，AES-128 的密文區塊鏈接模式、計數器模式、帶密文區塊鏈接-訊息驗證碼計數器模式、以及 Galois 計數器模式的正規化 Gini 不純度都相對較高；此外，這四種模式之間的正規化 Gini 不純度沒有顯著差異，表明它們是相對安全的模式。

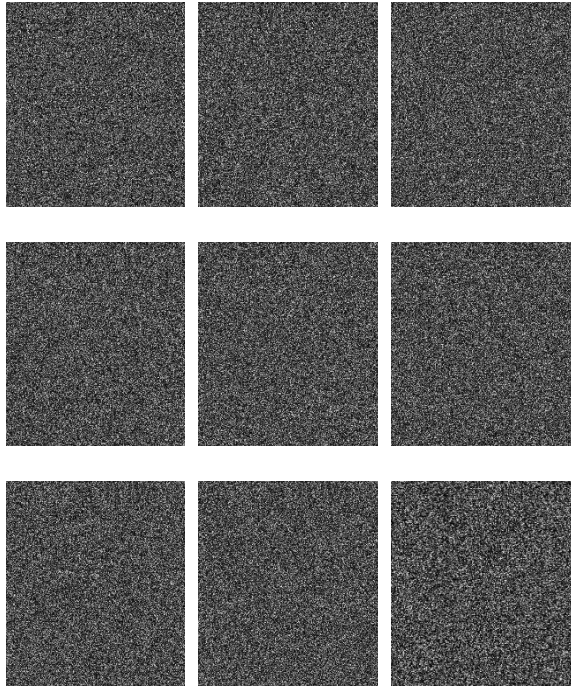

圖 3 AES-128密文區塊鏈接模式產製的加密影像

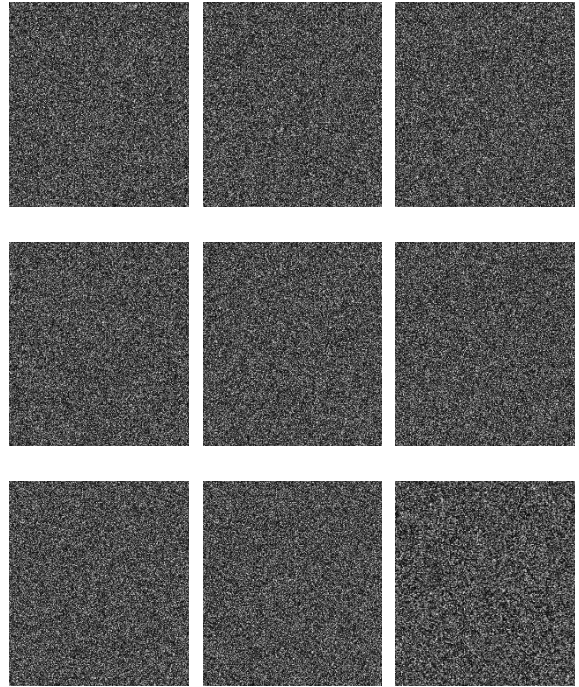

圖 4 AES-128計數器模式產製的加密影像

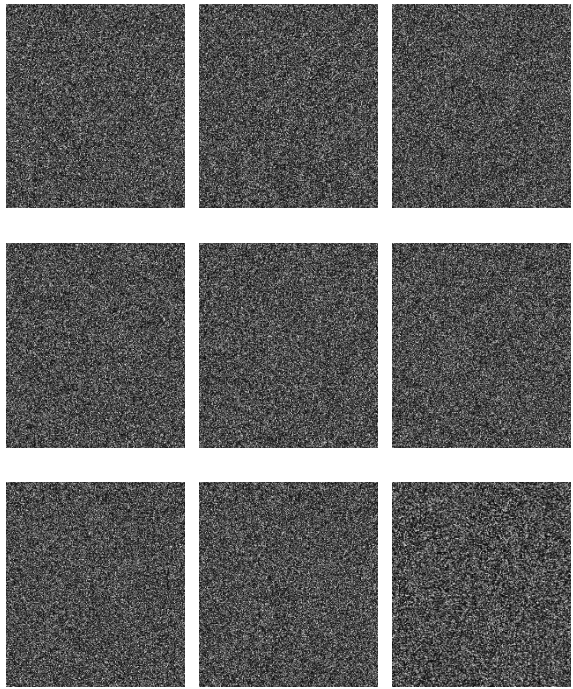

圖 5 AES-128帶密文區塊鏈接-訊息驗證碼計數器模式產製的加密影像

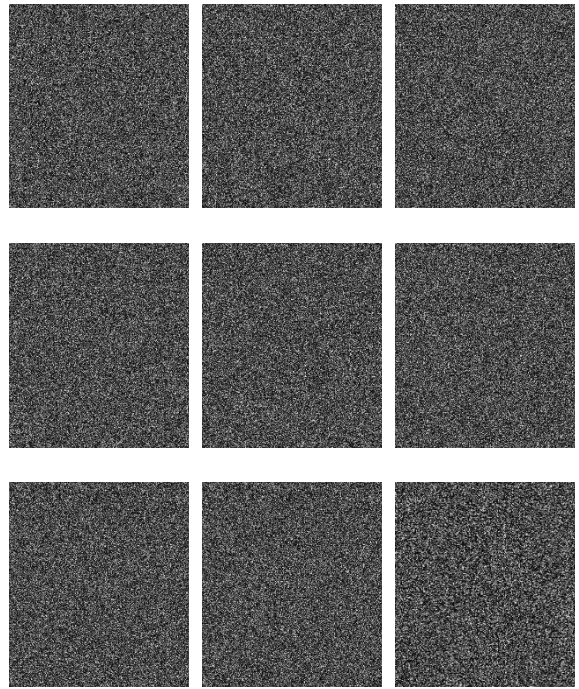

圖 6 AES-128 Galois 計數器模式產製的加密影像

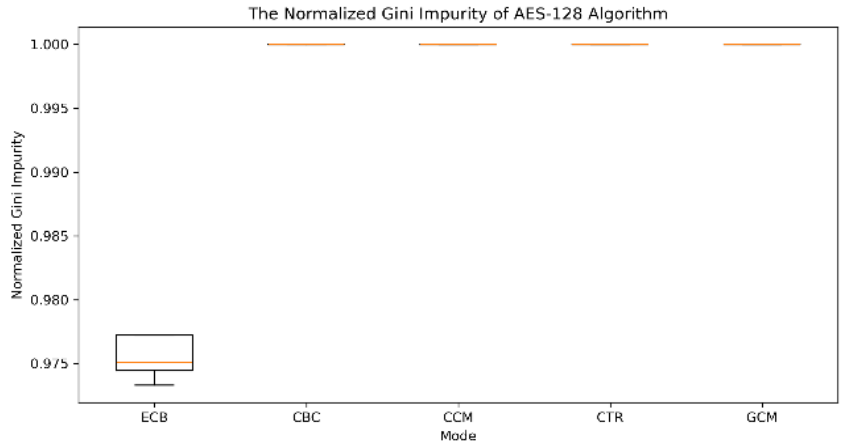

圖 7 AES-128各種模式的正規化 Gini 不純度

### 3.2 AES-192實驗結果

圖8是使用 AES-192電子密碼本模式產製的加密影像。從實驗結果可以觀察到，儘管秘密金鑰長度增加，但在電子密碼本模式的計算結果中仍可辨識出加密影像與原始影像的相似之處。

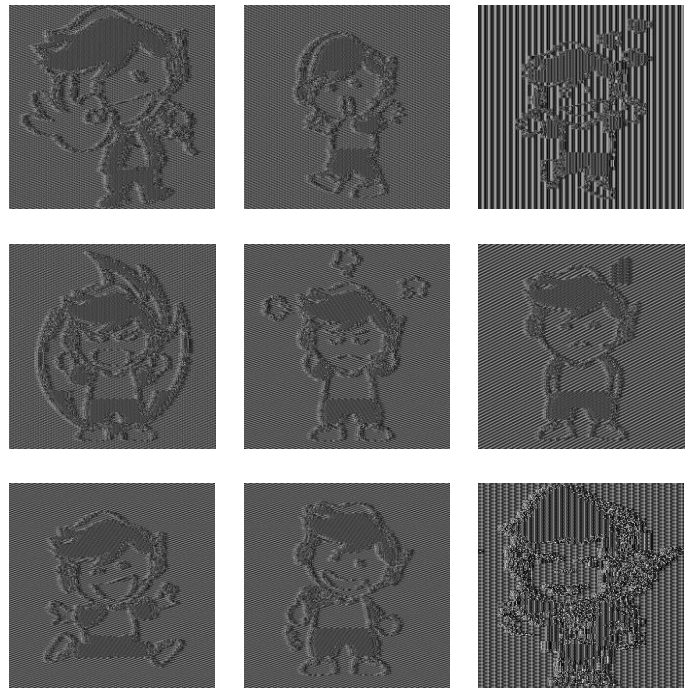

圖 8 AES-192電子密碼本模式產製的加密影像

由於論文篇幅限制，本小節未呈現基於 AES-192密文區塊鏈接模式、計數器模式、帶密文區塊鏈接-訊息驗證碼計數器模式、以及 Galois 計數器模式產製的加密影像；然而，這四種模式產製的加密影像與原始影像具有顯著差異。圖9比較每種模式的安全性，實驗結果指出電子密碼本模式的正規化 Gini 不純度最低，表示電子密碼本模式的安全性最低。

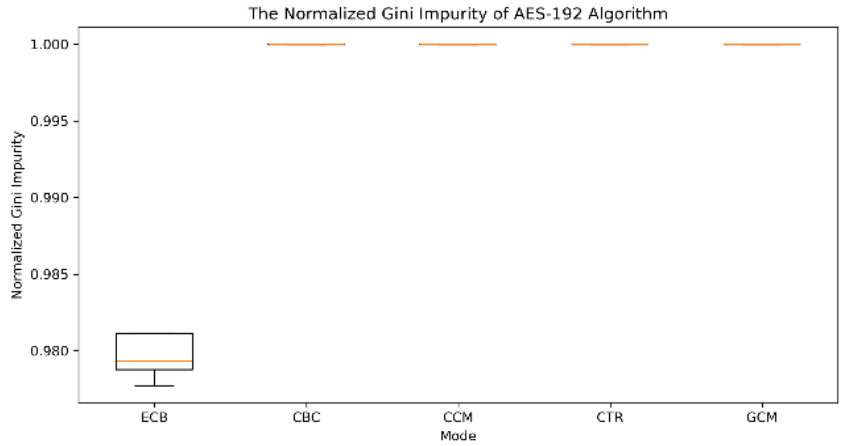

圖 9 AES-192各種模式的正規化 Gini 不純度

### 3.3 AES-256實驗結果

圖10顯示由 AES-256電子密碼本模式產製的加密影像。從實驗結果中可以觀察到，加密影像與原始影像的內容仍存在相似之處。因此，隨著秘密金鑰長度的增加，安全性並沒有顯著提高。所以 AES-256電子密碼本模式是不安全的。

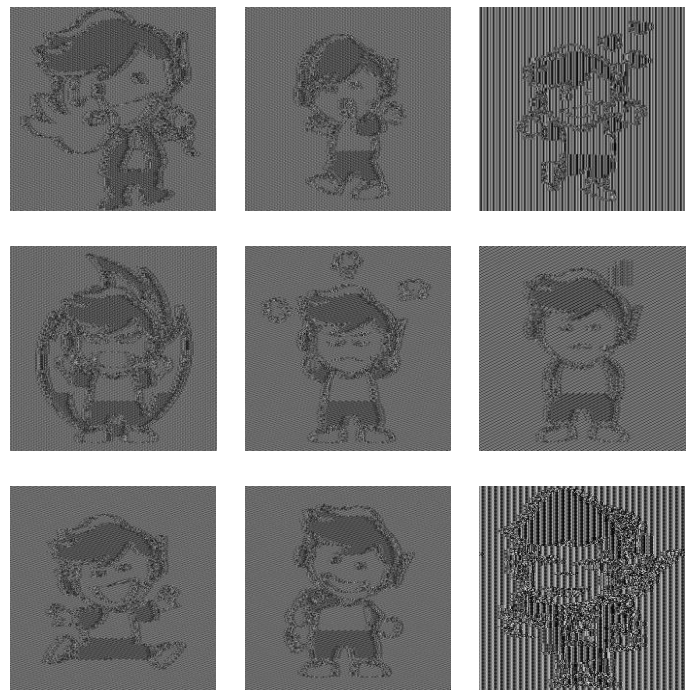

圖 10 AES-256電子密碼本模式產製的加密影像

由於論文篇幅限制，本文未提供由 AES-256密文區塊鏈接模式、計數器模式、帶密文區塊鏈接-訊息驗證碼計數器模式、以及 Galois 計數器模式產製的加密影像；然而，實驗結果表明，這四種模式產製的加密影像與原始影像之間存在顯著差異。觀察不到加密影像與原始影像的相似之處，表明這四種模式能夠提供安全的加密。

圖11使用本研究提出的正規化 Gini 不純度進行驗證。實驗結果表明，AES-256電子密碼本模式的正規化 Gini 不純度最低，表示安全性最低。

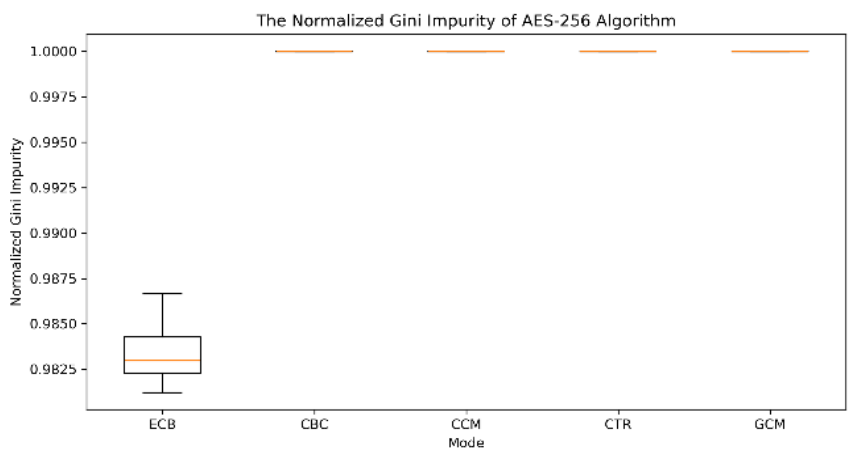

圖 11 AES-256各種模式的正規化 Gini 不純度

## 4. 小結與討論

本小節總結圖 7、圖 9 和圖 11 的實驗結果，以進行比較 AES-128、AES-192 和 AES-256 各種模式的正規化 Gini 不純度，如表 1 所示。根據實驗結果，AES-128、AES-192 和 AES-256 密文區塊鏈接模式、計數器模式、帶密文區塊鏈接-訊息驗證碼計數器模式、以及 Galois 計數器模式的正規化 Gini 不純度均接近於 0.99999，表示加密影像跟原始影像具有顯著差異，並確保這些模式具有足夠的安全性。然而，AES-128、AES-192 和 AES-256 電子密碼本模式的正規化 Gini 不純度皆低於其他模式，並且即使隨著秘密金鑰長度的增加而稍微增加，但並未顯著改善。此外，電子密碼本模式產製的加密影像中仍然存在與原始影像內容的相似之處，表明電子密碼本模式較不具備安全性。

表 1 AES-128、AES-192和 AES-256各種模式的正規化 Gini 不純度

| 模式 | AES-128 | AES-192 | AES-256 |
|---|---|---|---|
| ECB | 0.97769 | 0.98155 | 0.98457 |
| CBC | 0.99999 | 0.99999 | 0.99999 |
| CCM | 0.99999 | 0.99999 | 0.99999 |
| CTR | 0.99999 | 0.99999 | 0.99999 |
| GCM | 0.99999 | 0.99999 | 0.99999 |

## 四、計算時間評估與比較

在密碼學中，以安全性為第一優先，在滿足安全需求且同等安全等級下，執行效率才會成為評估因子之一。因此，本節分別比較 AES-128、AES-192 和 AES-256 電子密碼本模式、密文

區塊鏈接模式、計數器模式、帶密文區塊鏈接-訊息驗證碼計數器模式、以及 Galois 計數器模式的加密時間和解密時間。實驗環境的詳細信息已在第三.2 小節中介紹。在第四.1 小節將比較加密時間，而第四.2 小節將比較解密時間。

1. 加密時間比較結果

圖 12、圖 13 和圖 14 分別描述 AES-128、AES-192 和 AES-256 各種模式的加密時間。此外，表 2 總結 AES-128、AES-192 和 AES-256 各種模式的平均加密時間。

從實驗結果可以看出，由於電子密碼本模式將每個明文區塊直接加密成密文區塊，並且能夠對每個區塊進行平行計算，所以可以具備較短的加密時間。然而，儘管它效率高，電子密碼本模式是不安全的，因此不適合使用。

密文區塊鏈接模式和計數器模式都不需要產製訊息驗證碼，所以與帶密文區塊鏈接-訊息驗證碼計數器模式和 Galois 計數器模式相比，加密時間較短。然而，由於缺乏訊息驗證碼，密文區塊鏈接模式和計數器模式無法確保訊息的完整性。

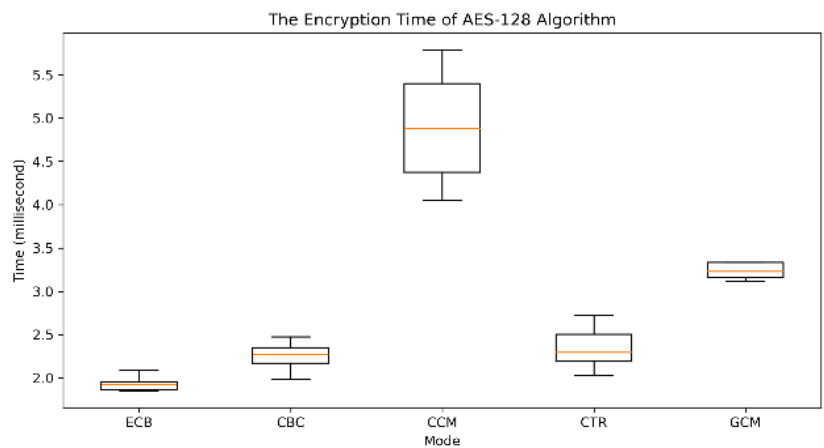

圖 12 AES-128各種模式的加密時間比較

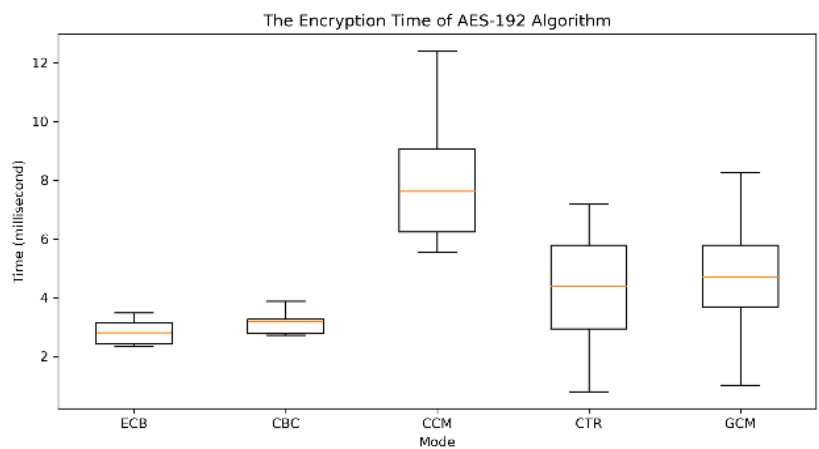

圖 13 AES-192各種模式的加密時間比較

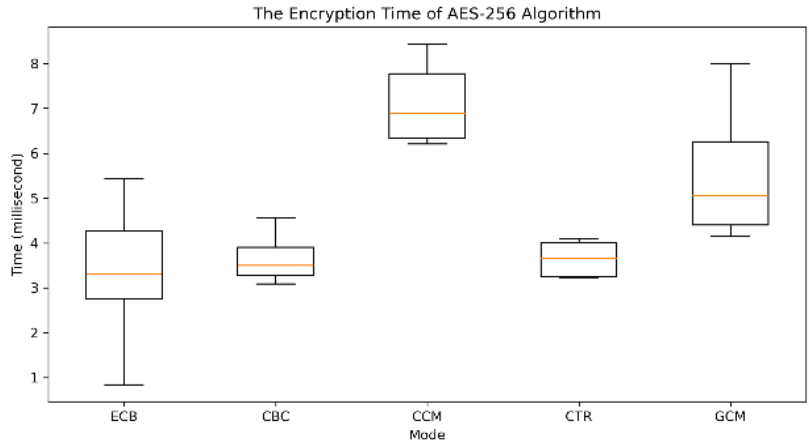

圖 14 AES-256各種模式的加密時間比較

表 2 AES-128、AES-192和 AES-256各種模式的加密時間比較

| 模式 | AES-128 | AES-192 | AES-256 |
|---|---|---|---|
| ECB | 2.60609 | 3.93449 | 4.88844 |
| CBC | 2.72313 | 3.60474 | 3.72583 |
| CCM | 5.27506 | 8.04113 | 7.08711 |
| CTR | 2.68617 | 4.39010 | 3.73377 |
| GCM | 3.86043 | 5.27984 | 5.63162 |

　　帶密文區塊鏈接-訊息驗證碼計數器模式在產製訊息驗證碼過程中需計算密文區塊鏈接，計算每個區塊時需等待前一個區塊計算完成，導致計算時間較長。另一方面，雖然 Galois 計數器模式也涉及訊息驗證碼計算，但它使用計數器模式計算每個密文區塊，然後在密文區塊上進行簡單的操作(如 XOR 運算符的訊息驗證碼計算)，所以 Galois 計數器模式的效率高於帶密文區塊鏈接-訊息驗證碼計數器模式。

## 2. 解密時間比較結果

　　圖 15、圖 16 和圖 17 分別描述 AES-128、AES-192 和 AES-256 各種模式的解密時間。此外，表 3 總結 AES-128、AES-192 和 AES-256 各種模式的平均解密時間。

　　由於解密過程與加密過程類似，把每個密文區塊都逐個解密為一個明文區塊。因此，解密過程中的計算工作量大致和加密過程中的計算工作量。在各種模式中，電子密碼本模式可以直接將每個密文區塊的解密平行轉換為一個明文區塊，所以具有較高的解密效率。然而，正如前面提到的，電子密碼本模式是不安全的。此外，解密效率從最高到最低排序如下：密文區塊鏈接模式、計數器模式、Galois 計數器模式和帶密文區塊鏈接-訊息驗證碼計數器模式。帶密文區塊鏈接-訊息驗證碼計數器模式由於需要在產製-訊息驗證碼過程中涉及密文區塊鏈接模式和計數器模式計算，所以導致較長的解密時間。

　　此外，實驗結果表明，隨著秘密金鑰長度的增加，解密時間也會增加。因此，儘管高級加密標準方法密碼學可以通過增加秘密金鑰長度來抵抗量子計算攻擊，但隨著秘密金鑰長度的延長，

計算時間將逐漸增加。

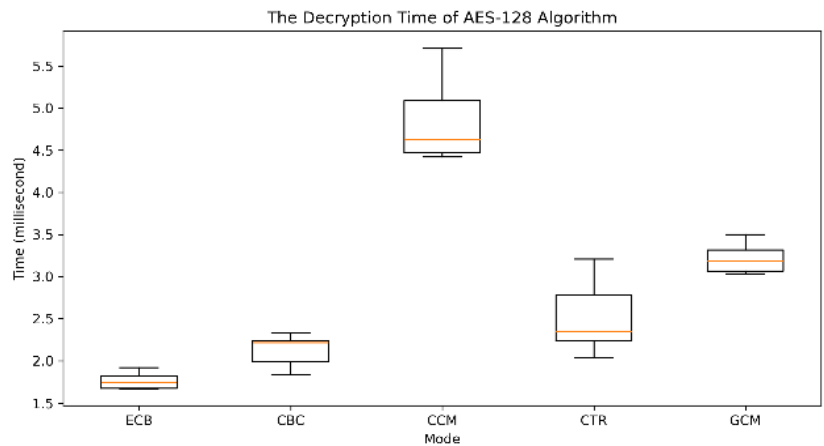

圖 15 AES-128各種模式的解密時間比較

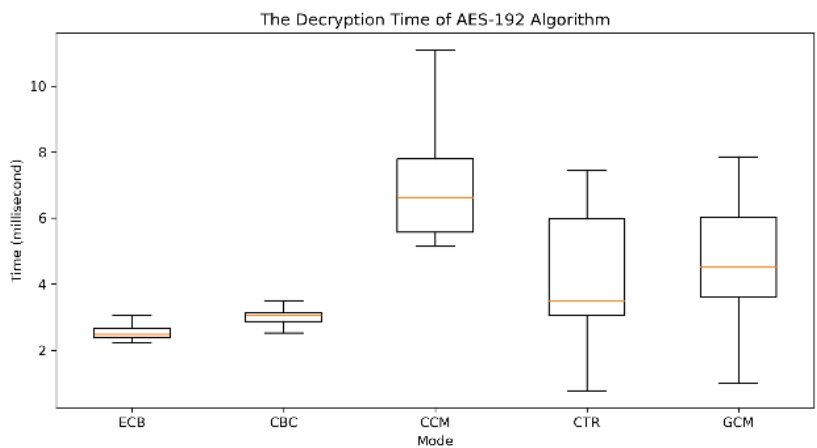

圖 16 AES-192各種模式的解密時間比較

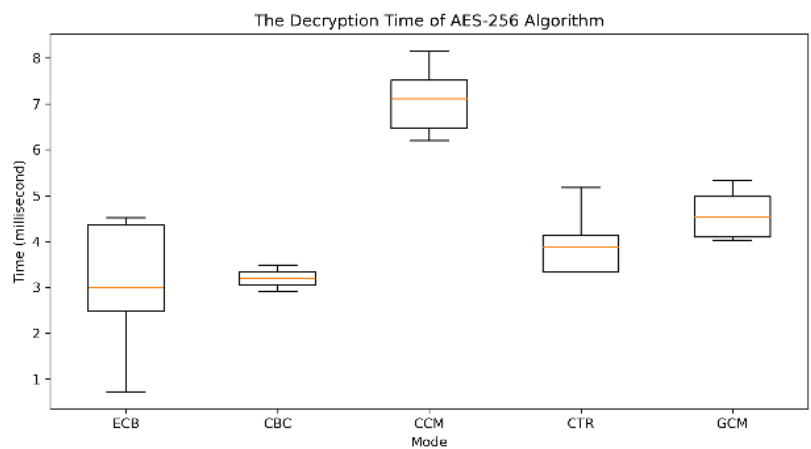

圖 17 AES-256各種模式的解密時間比較

表 3 AES-128、AES-192和 AES-256各種模式的解密時間比較

| 模式 | AES-128 | AES-192 | AES-256 |
|---|---|---|---|
| ECB | 2.39232 | 3.55077 | 4.76387 |
| CBC | 2.30042 | 3.37683 | 3.47647 |
| CCM | 4.98680 | 6.71384 | 7.16544 |
| CTR | 2.66613 | 4.08110 | 3.65854 |
| GCM | 3.79909 | 5.07418 | 5.69733 |

## 五、結論與未來研究

　　本研究主要提出正規化 Gini 不純度的安全性評估因子，用於評估 AES-128、AES-192 和 AES-256 各種模式(如電子密碼本模式、密文區塊鏈接模式、計數器模式、帶密文區塊鏈接-訊息驗證碼計數器模式、以及 Galois 計數器模式)。在實驗環境選擇數張原始影像，運用各種模式產製加密影像。實驗結果顯示，即使增加秘密金鑰長度，高級加密標準方法電子密碼本模式仍然是不安全的。然而，密文區塊鏈接模式、計數器模式、帶密文區塊鏈接-訊息驗證碼計數器模式、以及 Galois 計數器模式的安全性相對較高，其中帶密文區塊鏈接-訊息驗證碼計數器模式和 Galois 計數器模式進一步驗證訊息完整性。此外，實驗結果表明，Galois 計數器模式在加密和解密過程中都表現出較高的效率。因此，建議將來在系統部署可以考慮使用 Galois 計數器模式。

　　將來隨著量子計算的進步，雖然增加高級加密標準方法秘密金鑰長度可以抵抗量子計算攻擊，但隨著高級加密標準方法秘密金鑰長度的增加，計算時間也會增加。因此，未來設計高效且抵抗量子計算的對稱式密碼學方法及其模式，可以進一步提高對稱式密碼學的實用性。

## 致謝

## 參考文獻